\documentclass{elsart}
\usepackage[dvips]{graphicx}
\usepackage{amssymb}
\usepackage{array}
\usepackage{amsmath}

\begin{document}
\begin{frontmatter}

\title{Agent Simulation of Chain Bankruptcy}

\author[label1]{Yuichi Ikeda},
\author[label2]{Yoshi Fujiwara},
\author[label2]{Wataru Souma},
\author[label3]{Hideaki Aoyama},
\author[label4]{Hiroshi Iyetomi},

\address[label1]{Hitachi Research Institute, Tokyo 101-8010, Japan}
\address[label2]{Nict/ATR CIS Applied Network Science Lab., Kyoto 619-0288, Japan}
\address[label3]{Department of Physics, Kyoto University, Kyoto 606-8501, Japan}
\address[label4]{Department of Physics, Niigata University, Niigata 950-2181, Japan}

\begin{abstract}
We have conducted an agent-based simulation of chain bankruptcy. The propagation of credit risk on a network, i.e., chain bankruptcy, is the key to understanding large-sized bankruptcies.
In our model, decrease of revenue by the loss of accounts payable is modeled by an interaction term, and bankruptcy is defined as a capital deficit. Model parameters were estimated using financial data for 1,077 listed Japanese firms. 
Simulations of chain bankruptcy on the real transaction network consisting of those 1,077 firms were made with the estimated model parameters. Given an initial bankrupt firm, a list of chain bankrupt firms was obtained. This model can be used to detect high-risk links in a transaction network, for the management of chain bankruptcy.
\end{abstract}

\begin{keyword}
Econophysics \sep Agent-based simulation \sep Game theory \sep Firm dynamics \sep 
Complex network \sep Credit exposure management \sep Chain bankruptcy
\PACS 89.65.Gh 
\sep 02.50.Le 
\sep 89.75.Hc 
\end{keyword}
\end{frontmatter}

\section{Introduction}
\label{sec:intro}

\begin{figure}
\centerline{
\begin{minipage}{.45\linewidth}
\includegraphics[width=\linewidth, bb=0 0 331 229]{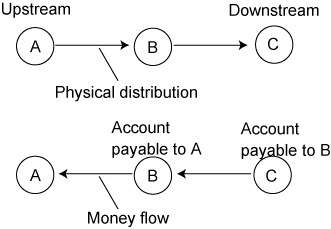}
\end{minipage}
}
\caption{\label{fig:BackGround}
A small transaction network consisting of three agents, A, B, and C.
If the downstream firm cannot pay its accounts payable due to bankruptcy, the upstream firm may go bankrupt. 
}
\end{figure}

A transaction network describes the physical distribution of products. 
Figure \ref{fig:BackGround} is a small transaction network consisting three agents, A, B, and C. The arrow indicates physical distribution. The left-hand side is upstream of the physical distribution.
The lower figure shows money flow. Note that the money flow arrow is in the opposite direction to that of physical distribution. 
The firm located upstream of physical distribution in a transaction network has the accounts receivable. 
On the other hand, the firm located downstream of physical distribution has the accounts payable. If the downstream firm cannot pay its accounts payable due to bankruptcy, the upstream firm may also go bankrupt. In this study bankruptcy is defined as a capital deficit. This propagation of credit risk on the transaction network is referred as "chain bankruptcy". 

An agent-based model of interacting firms, in which interacting firm agents rationally invest capital and labor in order to maximize payoff, has been developed and applied to some basic problems \cite{Ikeda2007a,Ikeda2007b}.
This paper describes the first attempt to apply the agent-based model to chain bankruptcy.
Although credit exposure management is an issue for individual firms \cite{Altman1968}, management for chain bankruptcy needs a knowledge of the whole transaction network. 
This paper is organized as follows. First, the propagation of credit risk is described.
Then, the agent-based model of chain bankruptcy is explained.
Finally, agent simulations are made for a real transaction network and the results are discussed.

\section{Propagation of Credit Risk}
\label{sec:CreditRisk}

\begin{figure}
\centerline{
\begin{minipage}{.5\linewidth}
\includegraphics[width=\linewidth, bb=0 0 259 184]{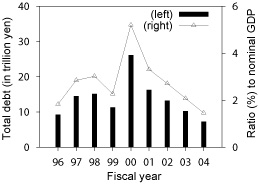}
\end{minipage}
}
\caption{\label{fig:Num_Debt_Gdp}
Temporal change in bankruptcy size.
Note that the total debt of bankrupt firms can be as large as a few percent of GDP.
}
\end{figure}

By using bankruptcy data for the last 10 years in Japan \cite{SMRJ},
it was shown that chain bankruptcies are by no means negligible \cite{Fujiwara2007}.
Temporal change of bankruptcy size is shown in Fig. \ref{fig:Num_Debt_Gdp}. 
Figure \ref{fig:Num_Debt_Gdp} shows total debt and its ratio to nominal GDP vs. fiscal year. 
From this figure, we see that the total debt of bankrupt firms can be as large as several percent of GDP.

\begin{figure}
\centerline{
\begin{minipage}{.45\linewidth}
\includegraphics[width=\linewidth, bb=0 0 223 181]{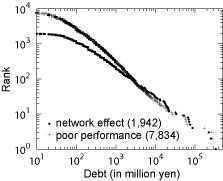}
\end{minipage}
}
\caption{\label{fig:d-cause}
The distribution of debt for bankrupt firms.
Note that the exponent for the distribution of bankruptcies due to the network effect is smaller than the distribution of bankruptcies due to the poor performance of individual firms. 
}
\end{figure}

Figure \ref{fig:d-cause} charts the distribution of debt for bankrupt firms.
Crosses indicate bankruptcies due to the poor performance of an individual firm.
Dots indicate bankruptcies due to the network effect, i.e., chain bankruptcy.
Both distributions are characterized by the power law distribution.
It should be noted that the exponent for the distribution of bankruptcies caused by the network effect is smaller than the distribution of bankruptcies caused by the poor performance of individual firms. This means that the chain bankruptcy effect is an important contributor to large-sized bankruptcies.

\section{Model of Chain Bankruptcy}
\label{sec:Model}

Revenue $R(t)$ of a $i^\textrm{\scriptsize th}$ firm agent is described by the time-evolution equation \cite{Ikeda2007a,Ikeda2007b}:  

\begin{equation}
R_i(t+1)=R_i(t) \left[ 
\frac{K_i^\textrm{\scriptsize (G)}(t+1)^{\alpha_i} L_i^\textrm{\scriptsize (G)}(t+1)^{\beta_i}}{K_i(t)^{\alpha_i} L_i(t)^{\beta_i}} 
+ \sum_{j \in Customers} f_{ij}+\sigma_i \epsilon_i \right],
\label{eq:EvEq}
\end{equation}

\begin{equation}
f_{ij}=k_{ij} \left( \frac{R_j(t)}{R_j(t-1)}-\frac{G(t)}{G(t-1)} \right).
\label{eq:Force}
\end{equation}

Here $K_i^\textrm{\scriptsize (G)}(t)$ and $L_i^\textrm{\scriptsize (G)}(t)$ are capital and labor, respectively \cite{Cobb1928}. 
Capital is the sum of tangible fixed assets and depreciation costs, thus capital is mainly investment in production facilities. 
On the other hand, labor is the sum of labor costs and staff costs. 
The suffix $(G)$ in the first term on the R.H.S. of Eq. (\ref{eq:EvEq}) indicates the solution of the game theory.
The second term on the R.H.S. of Eq. (\ref{eq:EvEq}) is an interaction term due to transactions. 
$G(t)$ of Eq. (\ref{eq:Force}) is gross domestic product. 
Material cost $C_i(t)$ is calculated using

\begin{equation}
C_i(t+1)=A_i K_i^\textrm{\scriptsize (G)}(t+1)^{\alpha_i}L_i^\textrm{\scriptsize (G)}(t+1)^{\beta_i}.
\end{equation}

The $i^\textrm{\scriptsize th}$ firm agent makes investment decisions for capital $K_i^\textrm{\scriptsize (G)}(t+1)$ and labor $L_i^\textrm{\scriptsize (G)}(t+1)$ in order to maximize profit $\Pi_i(t+1)$, defined by

\begin{equation}
\Pi_i(t+1)=R_i(t+1)-C_i(t+1)-r_i K_i^\textrm{\scriptsize (G)}(t+1)-L_i^\textrm{\scriptsize (G)}(t+1),
\end{equation}

under the investment decisions made by the rest of the agents. 
The agent's decision making was modeled using a genetic algorithm \cite{Ikeda2007b}.
This solution corresponds to the Nash equilibrium.
Model parameters $\alpha_i$, $\beta_i$, and $k_{ij}$ were empirically estimated as described later. 

\begin{figure}
\centerline{
\begin{minipage}{.55\linewidth}
\includegraphics[width=\linewidth, bb=0 0 134 110]{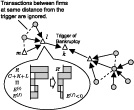}
\end{minipage}
}
\caption{\label{fig:CBlogic1}
The chain bankruptcy algorithm.
If equity at the end of the term is less then zero $E_l^{(f)}(t+1)<0$, 
the $l^\textrm{\scriptsize th}$ firm goes bankrupt (capital deficit) and cannot pay its accounts payable to the $m^\textrm{\scriptsize th}$ firm.
}
\end{figure}

 Figure \ref{fig:CBlogic1} explains the chain bankruptcy algorithm. Bankruptcy of the $k^\textrm{\scriptsize th}$ firm decreases revenue of the $l^\textrm{\scriptsize th}$ firm through the loss of accounts payable. Equity at the end of term $E_l^{(f)}(t+1)$ is calculated by adding operating profit to equity at the beginning of the term: 

\begin{equation}
E_l^{(f)}(t+1) = E_l^{(i)}(t+1) + \Pi_l(t+1),
\end{equation}

where the second term on the R.H.S. is equity at the beginning of the term. 

If equity at the end of the term is less then zero $E_l^{(f)}(t+1)<0$, 
the $l^\textrm{\scriptsize th}$ firm goes bankrupt (capital deficit) and cannot pay its accounts payable to the $m^\textrm{\scriptsize th}$ firm.

\begin{figure}
\centerline{
\begin{minipage}{.7\linewidth}
\includegraphics[width=\linewidth, bb=0 0 169 97]{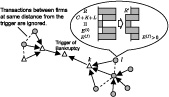}
\end{minipage}
}
\caption{\label{fig:CBlogic2}
If equity at the end of the term is larger then zero$E_l^{(f)}(t+1)>0$, 
chain bankruptcy stops at the lth firm.
This situation will occur when equity of the $l^\textrm{\scriptsize th}$ firm is large, or when the amount of the transaction (i.e., strength of interaction $k_{ij}$) is small.
}
\end{figure}

On the other hand, if equity at the end of the term is larger then zero $E_l^{(f)}(t+1)>0$, 
chain bankruptcy stops at the $m^\textrm{\scriptsize th}$ firm, as shown in Fig. \ref{fig:CBlogic2}. This situation will occur when equity of the $l^\textrm{\scriptsize th}$ firm is large, or when the amount of the transaction (i.e., the strength of interaction $k_{ij}$) is small.

The next step was calibration of the parameters. Parameters were estimated using financial data for 1,077 listed Japanese firms. Error was given by Eq. (\ref{eq:EvEq}), 

\begin{equation}
\frac{\epsilon_i}{\sigma_i} = \frac{1}{\sigma_i^2} \left[ \frac{R_i(t+1)}{R_i(t)}- \left( \frac{K_i(t+1)}{K_i(t)} \right)^{\alpha_i} \left( \frac{L_i(t+1)}{L_i(t)} \right)^{\beta_i} - \sum_j f_{ij} \right]. 
\end{equation}

From previous data analysis \cite{Ikeda2008}, it is known that the error distribution is normal. Thus, the likelihood $f(R, K, L, GDP | \alpha, \beta, k)$ is defined by 

\begin{equation}
f(R, K, L, GDP | \alpha, \beta, k)=(2\pi \sigma^2)^{-\frac{T}{2}} \prod_{t=1}^{T} \exp \left( -\frac{\epsilon_i^2}{2\sigma_i^2} \right),
\label{eq:Likelihood}
\end{equation}

and the log-likelihood $l(R, K, L, GDP | \alpha, \beta, k)$ is written by taking the logarithm of Eq. (\ref{eq:Likelihood})

\begin{equation}
l(R, K, L, GDP | \alpha, \beta, k)= \ln (2\pi \sigma^2)^{-\frac{T}{2}} - \sum_{t=1}^{T} \frac{\epsilon_i^2}{2\sigma_i^2}.
\label{eq:Loglikelihood}
\end{equation}

\begin{figure}
\centerline{
\begin{minipage}{.8\linewidth}
\includegraphics[width=\linewidth, bb=0 0 659 479]{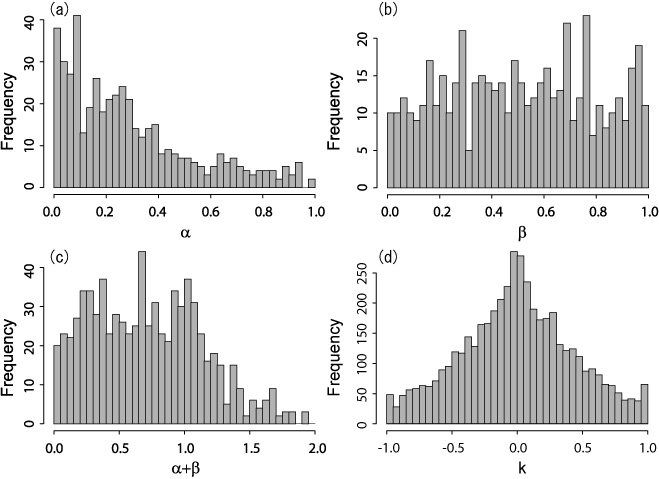}
\end{minipage}
}
\caption{\label{fig:Parameters}
Histograms of obtained parameters.
Model parameters were estimated with financial data using the maximum-likelihood method for 1,077 listed Japanese firms. Estimation errors were within several percent for most firms. 
}
\end{figure}

\begin{figure}
\centerline{
\begin{minipage}{.8\linewidth}
\includegraphics[width=\linewidth, bb=0 0 634 471]{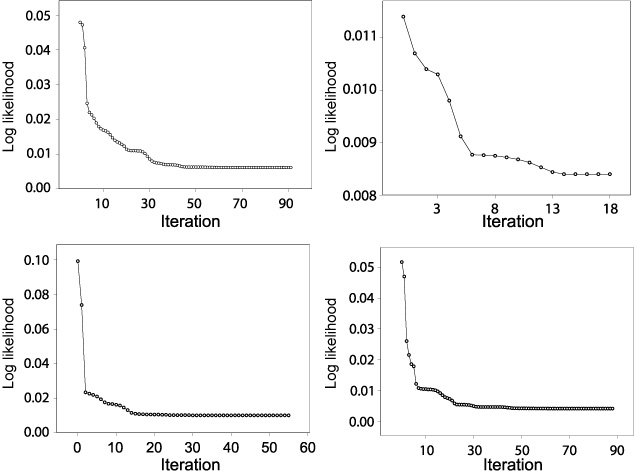}
\end{minipage}
}
\caption{\label{fig:Loglikelihood}
Convergence of the 2nd term of log likelihood for four arbitrarily chosen firms.
}
\end{figure}

\begin{figure}
\centerline{
\begin{minipage}{.6\linewidth}
\includegraphics[width=\linewidth, bb=0 0 530 410]{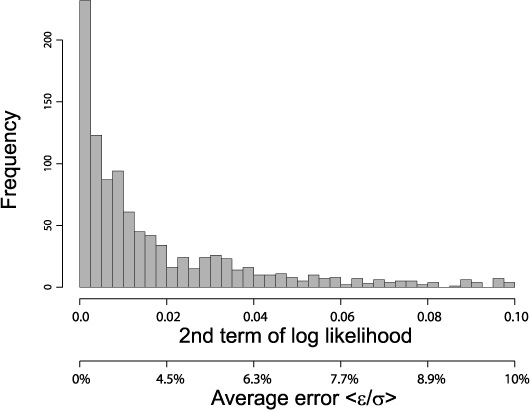}
\end{minipage}
}
\caption{\label{fig:Func}
Distribution of average error.
Estimation errors were within several percent for most firms. 
}
\end{figure}

Here $t=1$ to $T$ represent Japanese Fiscal Years (JFY) 1993 to JFY2003. 
We minimized the second term of the log-likelihood using the quasi-Newton method.  
Figure \ref{fig:Parameters} shows histograms of the obtained parameters 
Figures \ref{fig:Parameters} (a) to (d) are frequencies of $\alpha_i$, $\beta_i$, $\alpha_i+\beta_i$, and $k_{ij}$, respectively.
Figure \ref{fig:Loglikelihood} shows the convergence of the second term of log likelihood for four arbitrarily chosen firms.
Finally, the distribution of average error is shown in Fig. \ref{fig:Func}. Average error is calculated from values of the second term of log likelihood. Estimation errors were within several percent for most firms.

\section{Agent Simulations}
\label{sec:Simulation}

Chain bankruptcy on a real transaction network, consisting of 1,077 listed Japanese firms, was simulated with the estimated parameters by creating bankruptcy at one firm, called the trigger firm. We created simulations for two cases. The trigger in the first case is firm A, belonging to the electrical machine industry sector, and the trigger in the second case is firm B, belonging to the chemical industry sector. 

\begin{figure}
\centerline{
\begin{minipage}{.7\linewidth}
\includegraphics[width=\linewidth, bb=0 0 435 440]{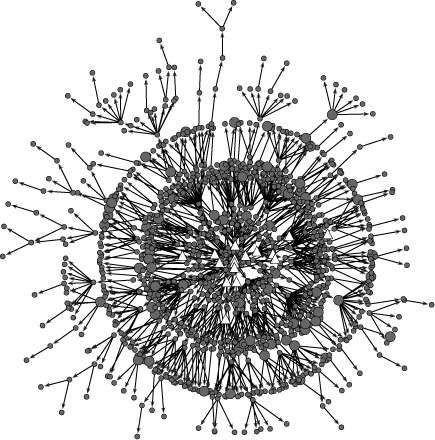}
\end{minipage}
}
\caption{\label{fig:Hitachi2}
Simulation result of chain bankruptcy triggered by a firm belonging to the electrical machine industry sector.
Bankrupt firms are indicated by white triangles. 
Arrows indicate the flow of money, which is in the opposite direction to physical distribution. 
}
\end{figure}

Figure \ref{fig:Hitachi2} is the chain bankruptcy simulation result triggered by firm A. Bankrupt firms are indicated by white triangles; firm A is located at the center of this network. Circles are suppliers of firm A; suppliers separated by many links are located on the outer circumference. Arrows indicate the flow of money, which is in the opposite direction to physical distribution. 
From this simulation, we obtained a list of chain bankrupt firms. 

\begin{figure}
\centerline{
\begin{minipage}{.7\linewidth}
\includegraphics[width=\linewidth, bb=0 0 470 454]{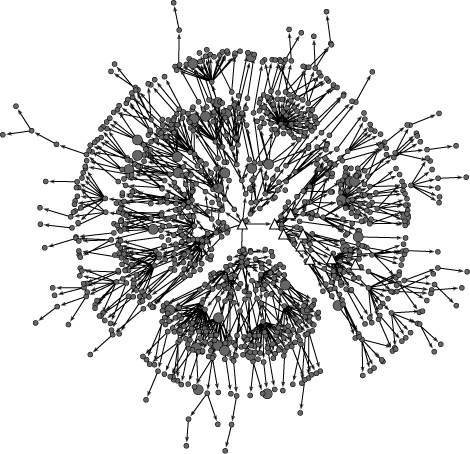}
\end{minipage}
}
\caption{\label{fig:HitachiChemical2}
Simulation result of chain bankruptcy triggered by a firm belonging to the chemical industry sector.
In this case, the bankruptcy chain is extended in only one direction, unlike the first case shown in Fig. \ref{fig:Hitachi2}.
}
\end{figure}

Figure \ref{fig:HitachiChemical2} is the chain bankruptcy simulation result triggered by firm B; firm B is located at the center of this network. In this case, the bankruptcy chain is extended in one direction only, unlike the first case.
Similarly, we obtained a list of chain bankrupt firms. The number of links is larger than in the first case, although the number of bankrupt firms is smaller.

\section{Conclusions}
\label{sec:Conclusions}

In this project we created agent-based simulations of the chain bankruptcy process. 
First, it was shown bankruptcies are an important economic phenomenon. The total debt of bankrupt firms can be as large as a few percent of GDP. Understanding the propagation of credit risk through a network, i.e., chain bankruptcy, is key to understanding large-sized bankruptcies.
Second, an agent model of chain bankruptcy was presented. In this model, decrease of revenue by the loss of accounts payable was modeled by an interaction term. Model parameters were estimated with financial data for 1,077 Japanese firms. 
Finally, simulations of chain bankruptcy on a real transaction network consisting of 1,077 listed Japanese firms were made with the estimated model parameters. 
This agent model will be useful for detecting high-risk links in the transaction network, for the management of chain bankruptcy.

Briefly, our plans for further study are as follows. 
The first task concerns the stochastic simulation of bankruptcy by taking into account the third term on the R.H.S. of Eq. (\ref{eq:EvEq}). A default correlation obtained from the stochastic simulation will be used as input data for Value at Risk simulation \cite{CreditMetrics1997}. The second task concerns a generalization of the agent model of chain bankruptcy. The current model focuses on a transaction network of manufacturers. The generalized model will be applicable to analyzing liquidity risk on a network of banks \cite{Cifuentes2005}.

\end{document}